\documentclass[aps,pra,twocolumn,superscriptaddress]{revtex4}
\usepackage{amssymb}
\usepackage{amsmath}
\usepackage{epsfig}
\usepackage{color}
\usepackage{graphics, graphicx}
\usepackage{bbold}
\usepackage{psfrag}
\usepackage{mathcomp}
\usepackage{subfigure}
\usepackage{verbatim}
\usepackage{color}
\usepackage[colorlinks,citecolor=blue]{hyperref}

\begin{document}

\title{Spin-tensor Meissner currents of ultracold bosonic gas in an optical lattice}
\author{Xiaofan Zhou}
\affiliation{State Key Laboratory of Quantum Optics and Quantum Optics Devices, Institute
of Laser spectroscopy, Shanxi University, Taiyuan 030006, China}
\affiliation{Collaborative Innovation Center of Extreme Optics, Shanxi University,
Taiyuan, Shanxi 030006, China}
\author{Suotang Jia}
\affiliation{State Key Laboratory of Quantum Optics and Quantum Optics Devices, Institute
of Laser spectroscopy, Shanxi University, Taiyuan 030006, China}
\affiliation{Collaborative Innovation Center of Extreme Optics, Shanxi University,
Taiyuan, Shanxi 030006, China}
\author{Xi-Wang Luo}
\email{luoxw@ustc.edu.cn}
\affiliation{CAS Key Laboratory of Quantum Information, University of Science and Technology of China, Hefei, Anhui 230026, China}
\affiliation{Synergetic Innovation Center of Quantum Information and Quantum Physics,
University of Science and Technology of China, Hefei, Anhui 230026, China}

\date{\today }

\begin{abstract}
We investigate the Meissner currents of interacting bosons subjected to a staggered artificial gauge field in a three-leg ribbon geometry, realized by spin-tensor--momentum coupled spin-1 atoms in a 1D optical lattice. By calculating the current distributions using the state-of-the-art density-matrix renormalization-group method, we find a rich phase diagram containing interesting Meissner and vortex phases, where the currents are mirror symmetric with respect to the {\color{red}middle leg} (i.e., they flow in the same direction on the two boundary legs opposite to that on the middle leg), leading to the spin-tensor type Meissner currents, which is very different from previously observed chiral edge currents under uniform gauge field. The currents are uniform along each leg in the Meissner phase and form vortex-antivortex pairs in the vortex phase. Besides, the system also support a polarized phase that spontaneously breaks the mirror symmetry, whose ground states are degenerate with currents either uniform or forming vortex-antivortex pairs. We also discuss the experimental schemes for probing these phases. Our work provides useful guidance to ongoing experimental research on synthetic flux ribbons and paves the way for exploring novel many-body phenomena therein.
\end{abstract}

\maketitle

\section{Introduction}
Charged particles in a magnetic field showcase a remarkable variety of macroscopic quantum phenomena, including quantized Hall resistance in topological insulators~\cite{Hasan2010,Qi2011}, Meissner effect in superconductors~\cite{Braunisch1992,Geim1998}.
Recent experimental advances in realizing synthetic gauge field in ultracold atomic system provide a powerful tool for exploring these novel phenomena
in a fully controllable, clean environment~\cite{Lin2011,Wang2012,Cheuk2012,Hamner2015,Wu2016,Huang2016,Li2017,SOCclock}.
Chiral Meissner (topological edge) currents have been observed experimentally with an atomic Bose (Fermi)
gas in both ladder~\cite{Boseladder1,Boseladder2} and three-leg ribbon geometries~\cite{Mancini2015,RamanRb}, where the two-dimensional (2D) lattice consists of the sites of an 1D optical lattice in the long direction and the internal atomic spin states forming a synthetic lattice in the short direction. Such synthetic dimensions enable investigating higher-dimensional physics beyond the physical dimensions of the systems and bring new opportunities in exploring novel quantum phenomena~\cite{Boada2012,SDprl14,Price2015}. The experimentally observed chiral currents result from an uniform gauge field that
is equivalent to ordinary spin-orbit coupling (i.e., spin-vector--momentum coupling) with the form
$q F_z$, where $q$ is the quasi-momentum along the optical lattice direction and $\mathbf{F}=\{F_x,F_y,F_z\}$ are the spin-vectors~\cite{Zhai2015}. Consequently, the corresponding Meissner currents also exhibit spin-vector properties, that is, atoms with opposite spins propagate along opposite directions~\cite{Boseladder2,Mancini2015,RamanRb,Livi2016}.

For higher spin ($\geq1$) systems (e.g., the synthetic three-leg ribbon),  it is well known
that there exist not only spin-vectors but also spin-tensors~\cite{tensor1,tensor2}. Spin-tensor--momentum coupling (STMC) in the form $q F_z^2$ has been proposed and experimentally realized recently~\cite{Luo2017,Li2020}, which can significantly modify the band structures (e.g., dark-state band, triply degenerate points) and lead to interesting many-body physics (e.g., magnetic stripe phase) in the presence of interactions~\cite{Luo2017,Mai2018,Hu2018,Lei2022,Zhang2022}. It was also shown that STMC in a 1D Mott bosonic lattice
can support interesting spin-tensor magnetism orders~\cite{zhou2020}.
So far, the studies of Meissner effects have been focused on the aforementioned spin-vector types with uniform synthetic gauge fields. Therefore, a natural and important question is to explore the Meissner effects of interacting atoms in optical lattices with STMC.

In this paper, we investigate the ground-state Meissner currents of spin-tensor--momentum coupled spin-1 bosons in a 1D optical lattice, using state-of-the-art density-matrix renormalization-group (DMRG) numerical methods~\cite{dmrg1,dmrg2}. The system corresponds to a synthetic three-leg ribbon with a staggered gauge field, where the magnetic domain wall is given by the middle leg. We are interested in the Meissner effects and distinguish different phases by examining the current and momentum distributions. In the non-interaction limit, there are two phases depending on the minima of the lowest single-particle band. The Meissner (M) phase with a single band minimum occurs when the inter-leg couplings are strong, where the 
Meissner current is uniform along each leg with its amplitude determined by the gauge field strength. As the inter-leg coupling decreases across some critical value, the system undergoes a transition to the vortex (V) phase with double band minima, where atoms can occupy both minima and their interference can lead to vortex structures of the currents, forming vortex-antivortex pairs. For both the M and V phases, the current distributions are always mirror symmetric with respect to the middle leg, they have a rank-2 spin-tensor form which is very different from the spin-vector Meissner current. In particular, here the current flows in the same direction on the two boundary legs, opposite to the current direction on the middle leg. Meanwhile, the rank-0 scalar charge current and rank-1 spin-vector current are both zero.

In the presence of interaction, the phase diagram is altered significantly. In the weak interaction region, ferromagnetic spin-spin interaction can dominate over density-density interaction and stabilize the V phase with equal populations on the two band minima. In the strong interaction region, the wave packet becomes broadened in the momentum space and the two wave packets at the two band minima would merge into one, and thus the system favors the M phase even when the lowest band has two minima. Interestingly, for intermediate interaction strength, the spin-spin interaction can induce a new polarized (P) phase that spontaneously breaks the mirror symmetry with respect to the middle leg, though the STMC (i.e., synthetic gauge field) preserves such symmetry. In the P phase, atoms start to occupy the dark {\color{red}middle} band induced by the STMC, and the ground-states are degenerate with currents either uniform or forming vortex-antivortex pairs. \textcolor{red}{The P phase resulting from the presence of dark middle band
is unique for the STMC systems, which is absent for spin-vector-coupled systems.} The Meissner currents in all phases persist no matter if the system belongs to a superfluid or Mott insulator.

\section{Model and Hamiltonian}

\begin{figure}[t]
\centering
\includegraphics[width = 8.5cm]{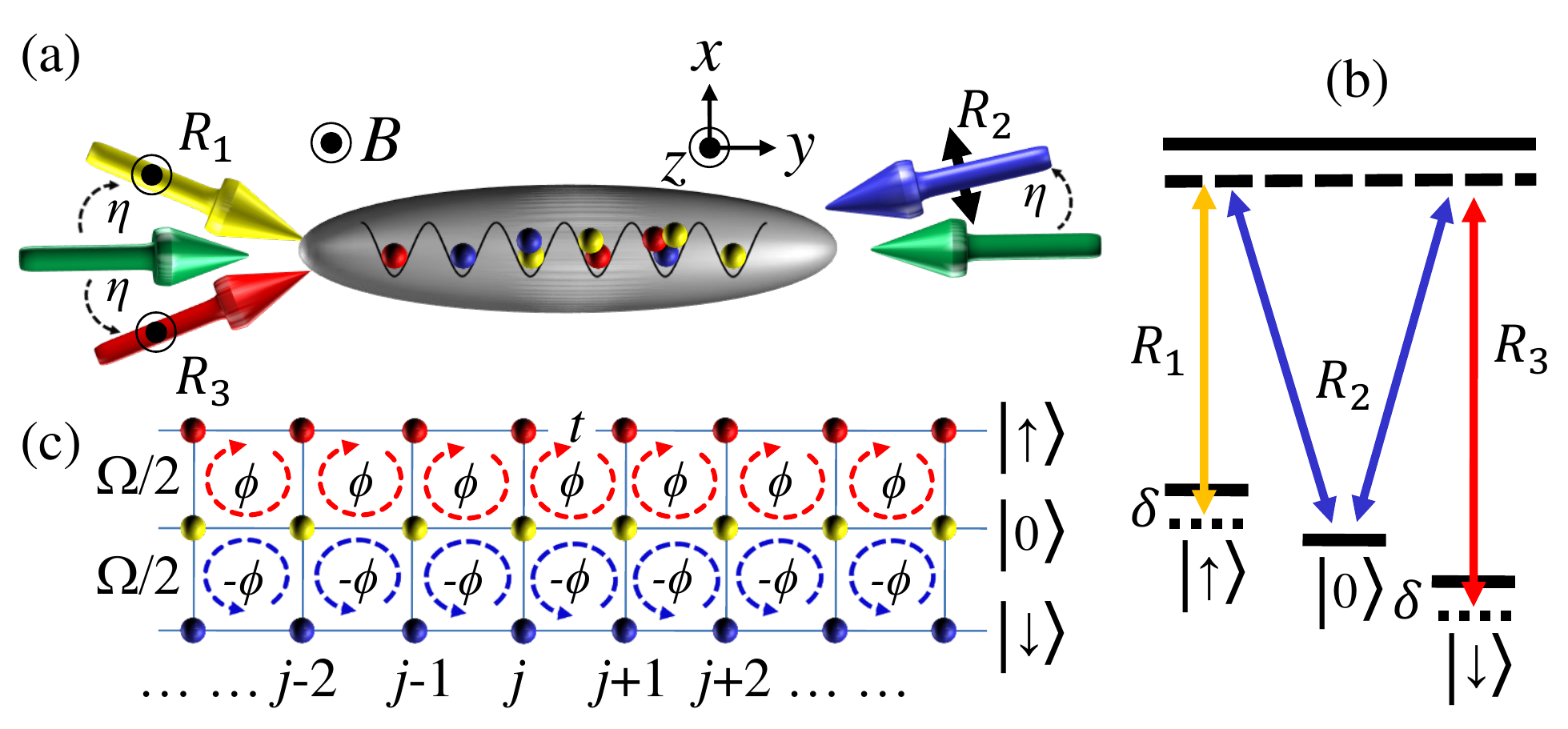} \vskip 0.0cm
\caption{(a) Schematics of the system setup. The ultracold atoms are trapped in a 1D optical lattice, which is generated by a pair of counterpropagating lasers with the wavelength $\protect\lambda _{\mathrm{L}}$ (green arrows). Three Raman lasers (yellow, blue and red arrows) with wavelength $\protect \lambda _{\Omega }$ and an angle $\protect\eta $ with respect to $y$-direction induce two Raman transitions. (b) Raman transitions between the
spin states $|0\rangle $ and $|\!\!\uparrow \!\!(\downarrow )\rangle$ with detuning $\delta$. (c) The effective three-leg ribbon with staggered gauge field along synthetic dimension.}
\label{experimental}
\end{figure}

\begin{figure*}[t]
\centering
\includegraphics[width = 12cm]{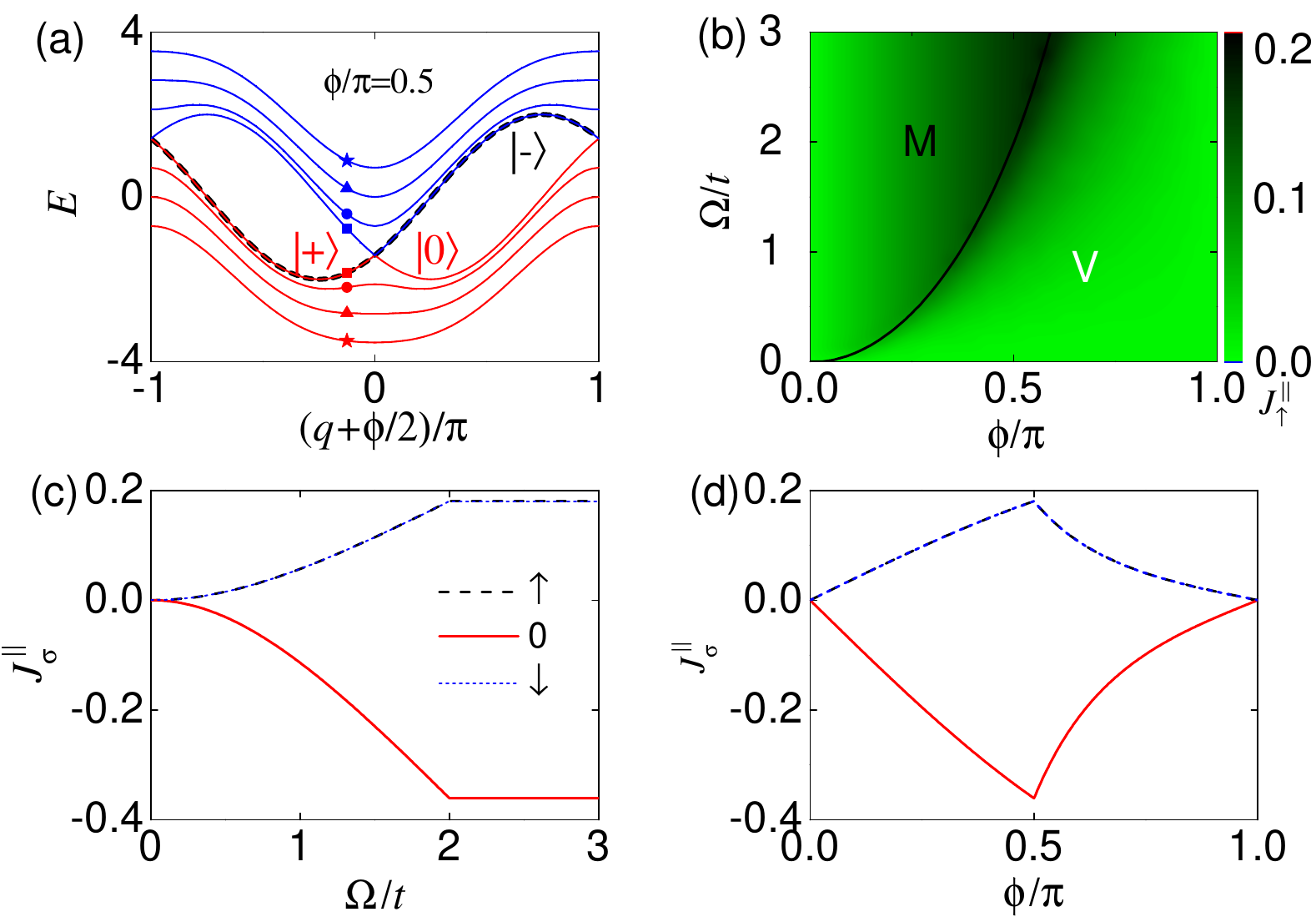} \hskip -0.05cm
\caption{{\color{red}(a) Single-particle band structure of
the Hamiltonian Eq.(\ref{Hk}) for different values of $\Omega/t$ with flux $\phi/\pi=0.5$ and $\Omega_c/t=2$.
The system exhibits three bands, including upper band (blue line), lowest band (red line) and the middle band (black dashed line). Blue (red) lines with different markers correspond to the upper (lowest) band with different $\Omega$, the square, dot, triangle, and star mark the bands with $\Omega/t=0,1,2,3$, respectively. The middle band is the dark band with spin state $| $-$ \rangle$, and is unaffected by $\Omega$.
The spin components $| 0 \rangle$ and including $|$+$\rangle$ are indicated around the corresponding band minima.
The lowest band has a single minimum for $\Omega>\Omega_c$ and two symmetric band minima for $\Omega<\Omega_c$.}
(b) The non-interacting phase diagram in the plane $\Omega \!\! - \!\! \phi$, with quantum phases vortex (V) and Meissner (M).
The color bar denotes the current $J_{\uparrow}^{\parallel}$, and the solid line is the critical line given by $\Omega_c$.
(c) (d) The profiles of the current  $J_{\sigma}^{\parallel}$ 
as functions of $\Omega$ and $\phi$, respectively, with $\phi/\pi=1/2$ in (c) and $\Omega/t=2.0$ in (d). The filling factor is $\rho = 1.0$ in (b)-(d).
We set $t$ as energy unit.}
\label{Noninteraction_phase_diagram}
\end{figure*}

We consider an experimental setup based on Bose-Einstein condensate (BEC) in a 1D optical lattice with STMC, as shown in Fig.~\ref{experimental}(a). A pair
of counter propagating laser with wavelength $\lambda_{\mathrm{L}}$ (green arrows) is used to realize the 1D optical lattices $V_{\mathrm{lat}}\left(
y\right) =-V_{0}\cos ^{2}\left( k_{\mathrm{L}}y\right) $ along the $y$-direction, with wave number $k_{\mathrm{L}}=2\pi /\lambda _{\mathrm{L}}$.
Three Raman lasers $R_{1}$ (yellow), $R_{2}$ (blue) and  $R_{3}$ (red) with wavelength $\lambda_{\Omega }$ and an incident angle $\eta $ with respect to lattice direction induce two Raman transitions between the spin states $|0\rangle $ and $|\!\!\uparrow\!\!(\downarrow )\rangle $ accompanied with the momentum transfer $2k_{\mathrm{R}}$~\cite{Li2020}, in which $k_{\mathrm{R}}=2\pi \cos (\eta )/\lambda _{\Omega }$, as shown in Fig.~\ref{experimental}(b). The tight-binding Hamiltonian reads
\begin{eqnarray}
H&=&-t\sum_{\left\langle j,j'\right\rangle,\sigma}\hat{%
b}_{j\sigma }^{\dag }\hat{b}_{j'\sigma }+ \frac{\Omega}{\sqrt{2}}
\sum_{j}\left( e^{i\phi j}\hat{b}_{j0}^{\dag }\hat{b}_{j+}+\mathrm{H.c.}%
\right)   \notag \\
&+&\sum_{j}\frac{U_{0}}{2}\hat{n}_{j}\left( \hat{n}_{j}-1\right) +\frac{U_{2}%
}{2}\left( \mathbf{S}_{j}^{2}-2\hat{n}_{j}\right) + \delta S_{j,z}^2,
\label{HTB}
\end{eqnarray}
where $\hat{b}_{j\sigma}^{\dag }$ ($\hat{b}_{j\sigma}$) is the Bose creating (annihilating) operator with spin basis $\sigma =\{\uparrow ,0,\downarrow \}$ and $\hat{b}_{j\pm}  = (\hat{b}_{j\uparrow} \pm \hat{b}_{j\downarrow})/\sqrt{2}$. The particle number operator is $\hat{n}_j = \sum_{\sigma} \hat{n}_{j\sigma}$ with $\hat{n}_{j\sigma} = \hat{b}_{j\sigma}^{\dag } \hat{b}_{j\sigma}$. $t$ is the tunneling amplitude between neighboring lattice sites, $\Omega $ is the Raman coupling strength. The flux $\phi =2\pi \cos (\eta)\lambda _{\mathrm{L}}/\lambda _{\Omega }=2\cos (\eta )k_{\mathrm{R}}a$
can be tuned by tuning the angle $\eta$. $\delta$ is the detuning for both $\left|\uparrow\right\rangle$ and $\left|\downarrow\right\rangle$ states.
{\color{red}For simplicity}, we set $\delta=0$ in the following. $U_{0}$ and $U_{2}$ are the density-density and spin-spin on-site interactions, which are related to the scattering lengths $a_{0,2}$ (corresponding to the channels with total spin 0 and 2 respectively) as $U_{0} = 4 \pi \hbar^2(a_0 + 2a_2)/3M$ and $U_{2} = 4 \pi \hbar^2 (a_2 - a_0)/3M$, where $M$ is the mass of the atom~\cite{Ho1988,Ohmi1998}. The spin dependent interaction for $^{23}$Na is antiferromagnetic with $U_{2}/U_{0}>0$, and ferromagnetic for $^{87}$Rb with $U_{2}/U_{0}<0$. $\mathbf{S}_{j}=\sum_{\sigma \acute{\sigma}}\hat{b}_{j\sigma }^{\dag }\mathbf{F}_{\sigma \acute{\sigma}} \hat{b}_{j\acute{\sigma}}$ is the total spin at site $j$, where $\mathbf{F}_{\sigma \acute{\sigma}}$
represent the matrix elements of the spin-vector operator. {\color{red}We also set $t$ as energy unit}.

\section{Phase diagram}
\subsection{Band structures}
We firstly discuss the single-particle band structure of the system. Effectively, the system corresponds to a three-leg ribbon with staggered gauge field as shown in Fig.~\ref{experimental}(c). The Hamiltonian without interaction can be written as after a unitary transformation
\begin{equation}
H_{0}=-te^{i\phi \gamma _{\sigma }}\!\!\!\!\!\sum_{<j,j'>,\sigma
}\!\!\!\hat{c}_{j\sigma }^{\dag }\hat{c}_{j'\sigma }+\frac{\Omega}{\sqrt{2}} \sum_{j} (
\hat{c}_{j0 }^{\dag }\hat{c}_{j+}+\mathrm{H.c.} )
\end{equation}%
where $(\hat{c}_{j\uparrow},\hat{c}_{j0},\hat{c}_{j\downarrow})^T=U (\hat{b}_{j\uparrow},\hat{b}_{j0},\hat{b}_{j\downarrow})^T$ with $U=\exp (i\phi jF_{z}^{2})$, and $\gamma _{\uparrow }=1$, $\gamma _{0}=0$, $\gamma _{\downarrow }=1$. Then, we Fourier transform the Hamiltonian and obtain
\begin{equation}
H_{0}=\sum_{q}\hat{C}_{q}^{\dag }\left[ -2t\cos
(qa+\phi F_{z}^{2})+\frac{\Omega}{\sqrt{2}} F_{x}\right] \hat{C}_{q},
\label{Hk}
\end{equation}%
where $\hat{C}_{q}=(\hat{c}_{q\uparrow},\hat{c}_{q0},\hat{c}_{q\downarrow})^T$ is the corresponding operator in quasi-momentum $q$ space, and $a=\pi /k_{\mathrm{L}}$ is the lattice constant, $F_{z}^{2}$ is the rank-2 spin tensor. We see that the flux $\phi$ now represent the strength of STMC. The system exhibits three bands after diagonalizing the Hamiltonian Eq.(\ref{Hk}), as shown in Fig.~\ref{Noninteraction_phase_diagram}(a).
The top and bottom bright-state bands exhibit the same behavior as a spin-orbit-coupled spin-1/2 system with spin states
$| 0 \rangle$ and $| $+$ \rangle=(\left|\uparrow\right\rangle+\left|\downarrow\right\rangle)/\sqrt{2}$ [see the color lines in Fig.~\ref{Noninteraction_phase_diagram}(a)]. The middle band $E_q = -2t\cos(q+\phi)$ is {\color{red}dark-state} band which always has the spin state $| $-$ \rangle=(\left|\uparrow\right\rangle-\left|\downarrow\right\rangle)/\sqrt{2}$ and is independent from $\Omega$ [see the black dash line in Fig.~\ref{Noninteraction_phase_diagram}(a)] since it is decoupled from the Raman lasers. However, the dark-state band plays an important role on both ground-state and Meissner current distributions in the presence of interactions.

\subsection{Order parameters}
\label{order parameters}
Before we calculate the phase diagram of the system, in the following, we first introduce some order parameters that we will use to distinguish different phases.

\emph{Currents.---}Since we are interested in the Meissner effects, the most important property here would be the current along the legs and the rungs of
the ribbon. Local averaged currents along the leg-direction can be define
as~\cite{Boseladder1,Boseladder2,helicalliquidnc,zhou2017current},
\begin{align}
J_{j, \sigma }^{\parallel } =it \langle \hat{b}_{j+1 \sigma
}^{\dag }\hat{b}_{j \sigma }-h.c.\rangle, J_{ \sigma }^{\parallel } =\frac{1}{L}\sum_{j}J_{j, \sigma
}^{\parallel }
\end{align}%
Similarly, we can define the rungs currents along the synthetic direction~\cite%
{Boseladder2,zhou2017current},
\begin{align}
J_{j,0 s}^{\perp } =i \Omega \langle e^{i\phi j}\hat{b}_{j 0 }^{\dag }\hat{b}%
_{j s}-h.c. \rangle
\end{align}%
with $s = \{ \uparrow, \downarrow \}$ .
We then introduce the current orders
\begin{align}
I_{0 s }^{\perp } =\frac{1}{L}\sum_{j}\left\vert J_{j,0 s }^{\perp
}\right\vert \text{ and }
I_{ z }^{\parallel } =\frac{1}{L}\sum_{j}|J_{j, \uparrow
}^{\parallel}-J_{j, \downarrow
}^{\parallel}|.
\end{align}%
The rungs current order $I_{0 s }^{\perp }$ distinguishes the M phase (with vanishing $I_{0 s }^{\perp }$) from the V phase (with finite $I_{0 s }^{\perp }$). The leg current order $I_{ z }^{\parallel }$ characterizes current polarization associated to the mirror symmetry with respect to the middle leg (a non-vanishing $I_{ z }^{\parallel }$ represent the breaking of the mirror symmetry).
Different phases can be identified by different current distributions and orders.

\emph{Spin moments.---}Experimentally, the spin moments of the atoms can be measured by spin-selective imaging of the density distributions
\begin{align}
n_{j,\sigma} = \langle \hat{b}_{j \sigma}^{\dag}\hat{b}_{j \sigma} \rangle.
\end{align}
The density distributions can be measured either in the basis $\sigma=\{\uparrow,0,\downarrow\}$ or the basis $\sigma=\{0,\pm\}$. We can define the spin moment orders as
\begin{align}
n_\sigma=\frac{1}{L}\sum_j n_{j,\sigma} 
\end{align}
which gives the population properties of the bands and can be used to characterize the phase transition. \textcolor{red}{We would like to mention that, $n_-$ corresponds to the occupation of the dark middle band.}

\emph{Entropy.---}The many-body interacting phase diagram can be further confirmed by entanglement entropy. Sharp features would emerge at the critical points in the von Neumann entropy defined as~\cite{Flammia2009,Hastings2010,Daley2012,Abanin2012,jiang2012,Islam2015,Amico2008}
\begin{align}
S_{\mathrm{vN}}=-\mathrm{Tr_{A}[\hat{\rho}_{A}\log \hat{\rho}_{A}]}
\end{align}
in which $\hat{%
\rho}_{A}=\mathrm{Tr_{B}|\psi \rangle \langle \psi |}$ is the reduced density matrix, and $|\psi \rangle$ is the ground-state wavefunction with $A,B$ corresponding to the left and
right half of the 1D chain.

\textcolor{red}{\emph{Low energy level spacing.---} The energy level spacing with respect to the ground state energy}, defined as
\begin{align}
\Delta_i = E_{i} - E_{0},
\end{align}
may also be used to signal the phase transition.
Where $E_{0}$ is the lowest eigenenergy, and the $E_{i}$ is the $i$-th order eigenenergy \textcolor{red}{with fixed particle numbers}.

\subsection{Noninteracting phase diagram}

In the noninteracting limit, the phase diagram is determined by the single-particle band structures. The lowest band presents a single minimum in the M phase $\Omega > \Omega_c$ and two symmetric minima in the V phase $\Omega < \Omega_c$, where $\Omega_c$ is the critical coupling strength, as shown in Fig.~\ref{Noninteraction_phase_diagram}(a). According to the number of minima, we can plot the single-particle phase diagram in the plane $\Omega \!\! - \!\! \phi$,
as shown in Fig.~\ref{Noninteraction_phase_diagram}(b), which is similar as the phase diagram of spin-1/2 SOC system~\cite{Boseladder2}.
However, the currents have a rank-2 spin-tensor form that are very different from the spin-vector Meissner current~\cite{Boseladder2,RamanRb}. In particular, here the current flows in the same direction on the boundary legs, opposite to the current direction on the middle leg. We calculate the ground-state currents of Hamiltonian Eq.(\ref{HTB}),  the averaged currents along each leg are shown in Figs.~\ref{Noninteraction_phase_diagram}(c) and \ref{Noninteraction_phase_diagram}(d). It is easy to check that the rank-0 scalar charge current $J^{\parallel }\equiv\sum_\sigma J_{ \sigma }^{\parallel }$
and rank-1 spin-vector current $J_{z}^{\parallel }\equiv\sum_\sigma \sigma J_{ \sigma }^{\parallel }$ are both zero,
while the rank-2 spin-tensor current $J_{zz}^{\parallel }\equiv\sum_\sigma \sigma^2 J_{ \sigma }^{\parallel }=2J_{\uparrow}^{\parallel }$ is  finite.
As we increase $\Omega$ for a fixed $\phi$, $J_{zz}^{\parallel }$ increases and has a saturated value above the critical point, as shown in Fig.~\ref{Noninteraction_phase_diagram}(c), the system undergoes a transition from V to M phase. When increasing the flux $\phi$ with $\Omega/t=2.0$, $J_{zz}^{\parallel }$ increase firstly and then decrease, reaching it maxima at the phase transition point from M to V phase, as shown in Fig.~\ref{Noninteraction_phase_diagram}(d).

\begin{figure}[t]
\centering
\includegraphics[width = 8.8cm]{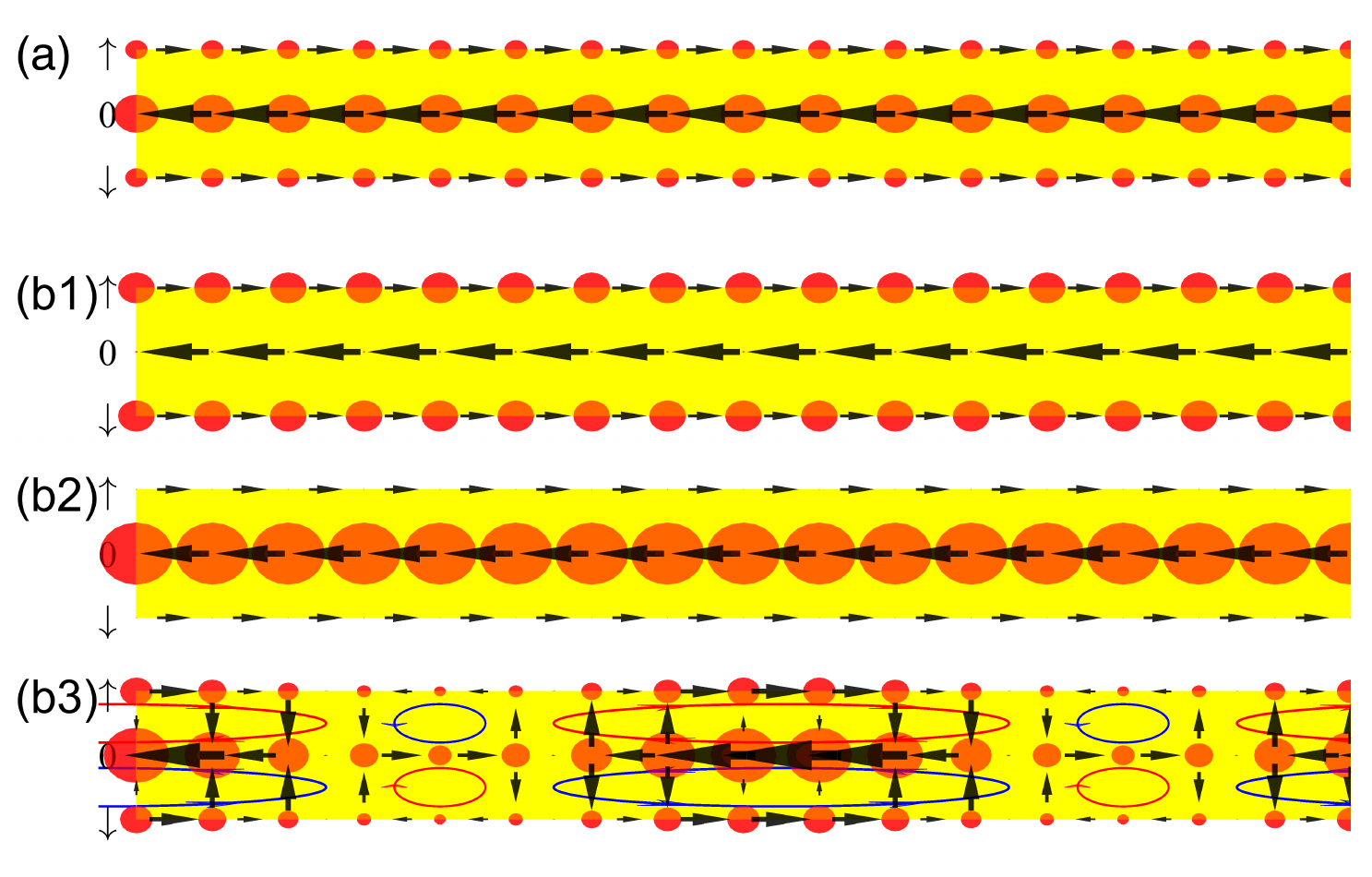} \hskip -0.05cm
\caption{The schematic of density distributions $ n_{j,\sigma} $, currents $J^{\parallel}_{j,\sigma}$ and $J^{\perp }_{j,0s}$ in different phases.
The points denote the density profile of the three-component with point size the amount of atoms. The current direction and strength is denoted by the arrows.
(a) The M phase with $\Omega/t = 2.5$ and $\phi/\pi=1/2$.
(b1)-(b3) The V phase with $\Omega/t = 0.2$ and $\phi/\pi=1/4$.
(b1) Left minimum ground-state [Blue dot in Fig.~\ref{Noninteraction_phase_diagram}(a)].
(b2) Right minimum ground-state [Red dot in Fig.~\ref{Noninteraction_phase_diagram}(a)].
(b3) Equal superposition of the two minima.}
\label{Noninteraction_phase}
\end{figure}

In order to clearly show the local current properties of the V and M phases, we plot the schematic of density distributions $ n_{j,\sigma} $, currents $J^{\parallel}_{j,\sigma}$ and $J^{\perp }_{j,0s}$, in Fig.~\ref{Noninteraction_phase}. The current is mirror symmetric with respect to the middle leg, analogous to the combination of spin-1/2 (two-leg ladder) vector current and it's mirror reflection. In the M phase, both the currents and the densities are uniform, as shown in Fig.~\ref{Noninteraction_phase}(a). In the V phase, the system exhibits two energy minima in the lowest band. Atoms may populate on either minimum with the same tensor current as shown in Figs.~\ref{Noninteraction_phase}(b1) and \ref{Noninteraction_phase}(b2), or a linear combination of the two minima with current vortex as shown in Fig.~\ref{Noninteraction_phase}(b3).

\begin{figure*}[tb]
\centering
\includegraphics[width = 12.0cm]{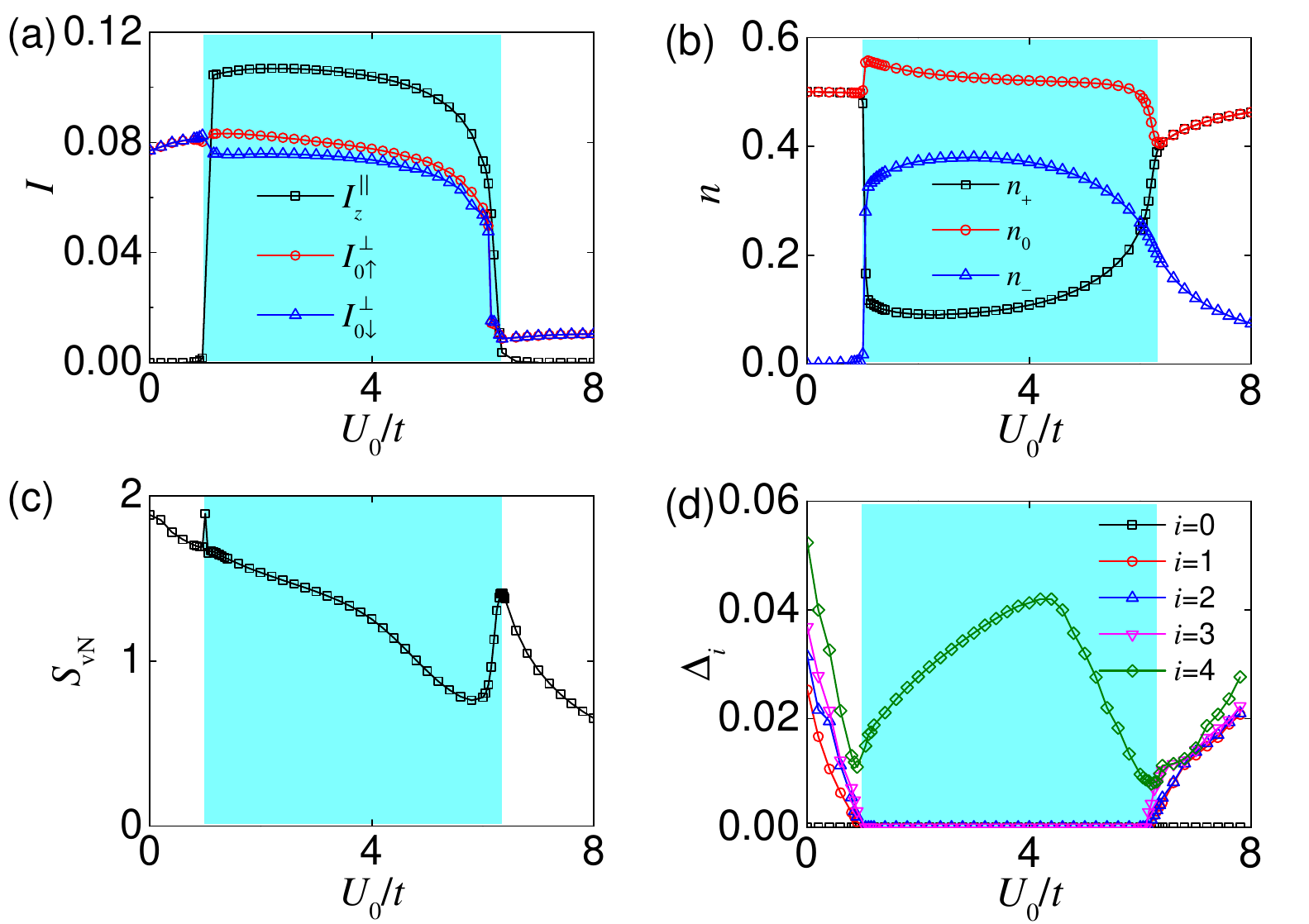} \hskip -0.0cm
\caption{(a) The current order parameters $I_{ z }^{\parallel }$, $I_{0 s }^{\perp }$ versus interaction strength $U_0$.
(b) Spin moment orders $n_{\pm,0}$ versus $U_0$.
(c) Entropy $S_{\text{vN}}$ and (d) Excited energy gap $\Delta_i$
as functions of interaction strength $U_0$.
In all subfigures, we have $\Omega/t=0.2$, $\phi/\pi=1/4$, $U_2=-0.2U_0$, $\rho = 1.0$ and $L=64$.}
\label{phase transition}
\end{figure*}

\subsection{Interacting phase diagram}

The interplay between single-particle band structure and interaction is fundamental to many areas of modern physics. The ability to engineer competing interactions between atoms through lattice depth or Feshbach resonances~\cite{Chin2010} makes cold atomic system an ideal platform to generate a rich variety of many-body phenomena~\cite{Raghu2008,Bloch2008,Mueller2017,Rachel2018,Junemann2017,zhou2017SPT,Leseleuc2019}. Here, we perform the state-of-the-art DMRG calculations to calculate the many-body ground-states of the system \textcolor{red}{under open boundary conditions}. 
{\color{red}In our numerical simulations, we set the cutoff of single-site atom number as $n_\text{cutoff}=4$. The effect of such a cutoff can be neglected for strong interaction, while in the weak interaction region, the phases and phase boundaries may be slightly affected by the cutoff. We find that $n_\text{cutoff}=4$ is enough to determine the phase boundaries (see Appendix A).}
We set lattice size up to $L=64$, for which we retain 300 truncated states per DMRG block and perform 20 sweeps with a maximum truncation error $\sim10^{-7}$.

We consider the repulsive density-density interaction with $U_0>0$ and ferromagnetic spin-spin interaction with $U_2<0$ as for $^{87}$Rb or $^7$Li atoms. In Fig.~\ref{phase transition}, we plot the order parameters as functions of the interacting strength $U_0/t$, with fixed $\Omega/t=0.2$ and $\phi/\pi=1/4$, from which, three phases (i.e., V, P and M) as well as their critical boundaries can be identified. For weak interactions, the system stays in V phase with a large rung current $I_{0 s }^{\perp }$; for strong interactions, the system favors the M phase with nearly vanishing $I_{0 s }^{\perp }$. We want to mention that, finite-size effects would induce a tiny averaged rungs current $I_{0 s }^{\perp }$ in the M phase, since the leg currents must form a closed loop at the boundary through rungs. In both the V and M phase, we have $I_{ z }^{\parallel }=0$ due to the mirror symmetry, though $J_{ \sigma }^{\parallel }\neq0$. For intermediate interaction strength, the P phase emerges with $I_{ z }^{\parallel }\neq0$, as shown in Fig.~\ref{phase transition}(a), and the mirror symmetry is spontaneously broken. The P phase may also support vortex currents, leading to a large rungs current order $I_{0 s }^{\perp }$. {Different phases have different spin occupations, therefore the spin moment orders also exhibit sharp transitions at the phase boundary, as shown in Fig.~\ref{phase transition}(b).} 

Besides the currents and spin moments, the von Neumann entropy $S_{\mathrm{vN}}$ as well as the low energy level spacing $\Delta_{i}$ can also signal the transitions. 
\textcolor{red}{When the entanglement entropy $S_{\mathrm{vN}}$, an analytic function of correlations, is not analytic at some point, it  must correspond to a quantum phase transition (as long as the definition of the entanglement entropy is analytic at that point). 
As demonstrated in Fig.~\ref{phase transition}(c), sharp features in the entropy emerge at the critical phase transition points ($S_{\mathrm{vN}}$ or its derivatives show discontinuity)~\cite{Amico2008}.
The degeneracy of the ground state can be seen from low energy level spacing which signals the spontaneous symmetry breaking across the phase transition.
As demonstrated in Fig.~\ref{phase transition}(d), $\Delta_{i}$ vanish for $i<4$ in the P phase due to ground-state degeneracy.}

\begin{figure}[b]
\centering
\includegraphics[width = 8.8cm]{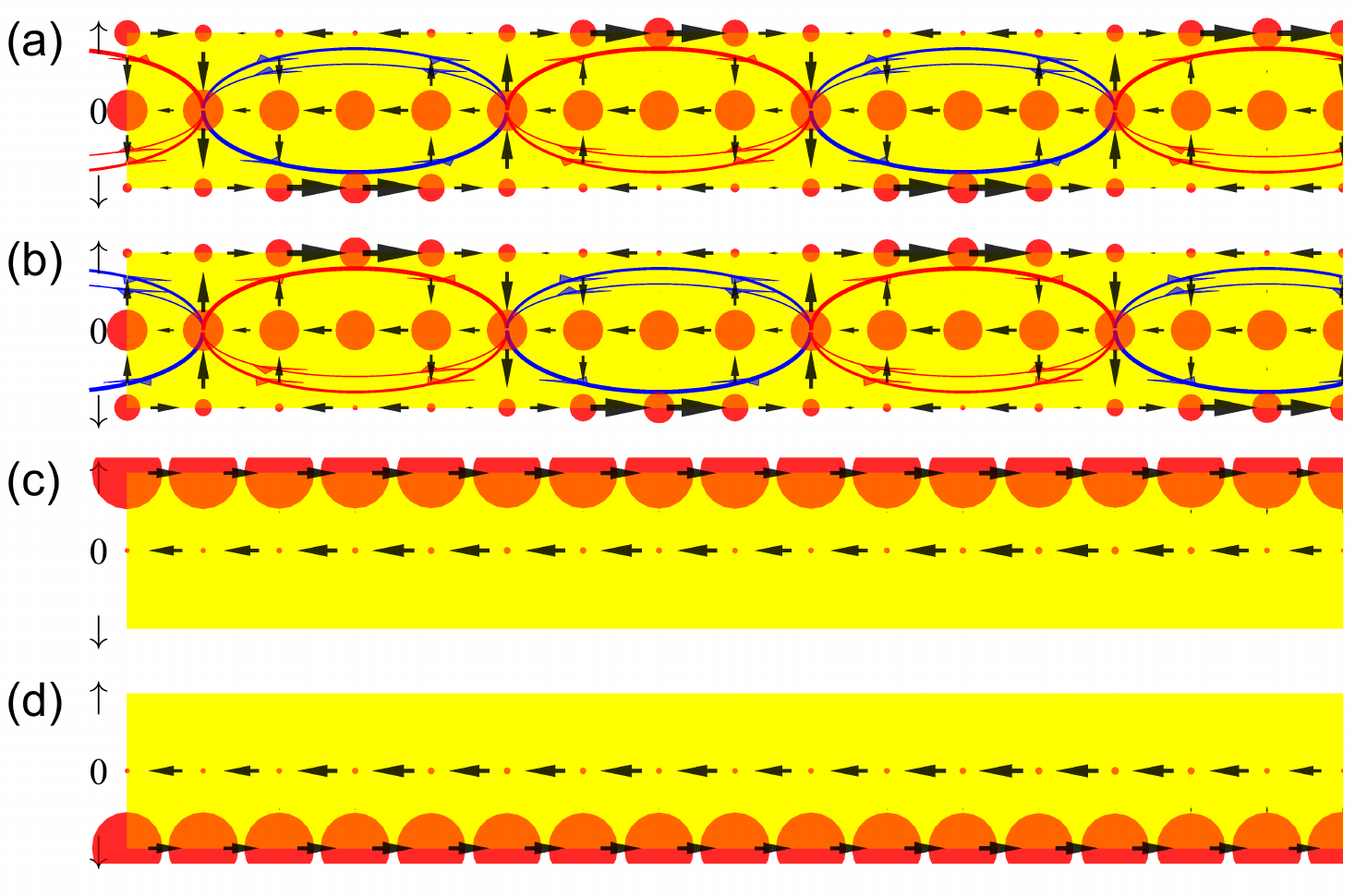} \hskip 0.0cm
\caption{(a)-(d) The schematic of density distributions $ n_{j,\sigma} $, currents $J^{\parallel}_{j,\sigma}$ and $J^{\perp }_{j,0s}$ of the four-fold degenerate ground-state in P phase, respectively.   In all subfigure, we have $\Omega/t = 0.2$, $\phi/\pi=1/4$, $U_0/t=4.0$, $U_2=-0.2U_0$, $\rho = 1.0$ and $L=64$ (only the central 17 sites are shown in the plot).}
\label{Vortex II}
\end{figure}

\begin{figure*}[t]
\centering
\includegraphics[width = 12.0cm]{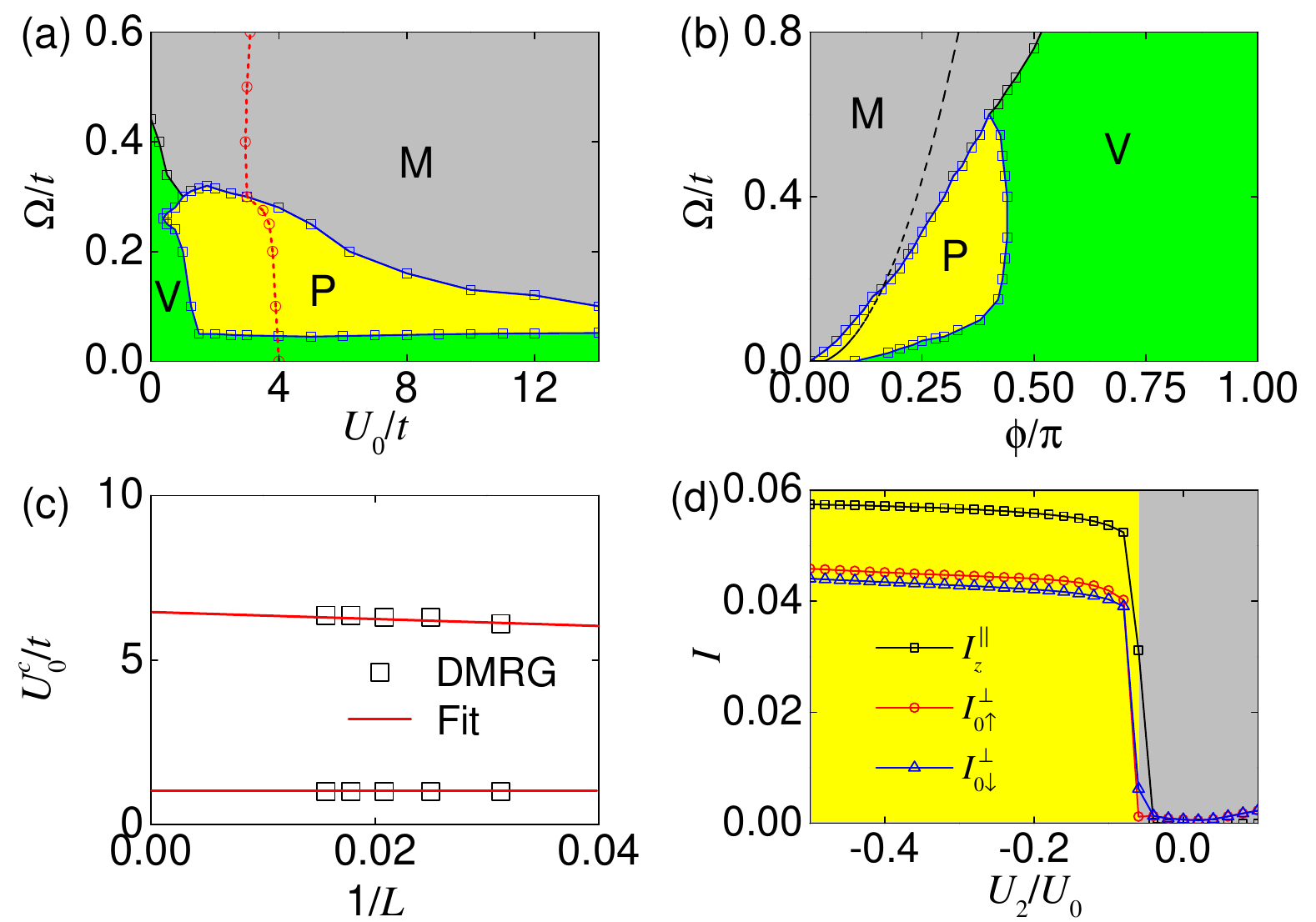} \hskip -0.05cm
\caption{The phase diagrams in the planes of (a) $\Omega \!\! - \!\! U_0$ with $\phi/\pi=1/4$,
and (b) $\Omega \!\! - \!\! \phi$ with $U_0/t=2$.
The red dashed line in (a) is the phase transition points between superfluid (left sides) and Mott insulator (right sides) phases, the black dashed line in (b) is the single-particle phase boundary.
(c) The finite-size scaling of the critical points $U^c_0/t$ of phase transitions,
with $\Omega/t=0.2$ and $\phi/\pi=1/4$.
In (a)-(c), $U_2=-0.2U_0$.
(d) The current order parameters $I_{ z }^{\parallel}$, $I_{0 s }^{\perp }$ versus interaction strength $U_2$, with $\Omega/t=0.2$, $\phi/\pi=1/4$, $U_0/t=2.0$ and $L=64$.
In all subfigure, we have $\rho = 1.0$.}
\label{phase diagram}
\end{figure*}

The interaction can drive the system from V (at weak interaction regime) to M (at strong interaction regime) phase though the lowest single-particle band has two minima. Different from the noninteracting V phase where the ground-state is an arbitrary linear combination of the two band minima, here the interacting V phase is stabilized by ferromagnetic spin-spin interaction, leading to an equal superposition of the two band minima. On one hand, the interference between two band minima would induce density modulation and increase density-density interaction energy. On the other hand, the superposition also generate spin orders which lower the ferromagnetic spin-spin interaction energy, and the system favors the V phase since the spin-spin interaction dominates over the density-density interaction. We want to mention that, for the V phase in the weak interacting region, the mean-field analysis predicts that the relative phase between atoms at two band minima can either be $0$ or $\pi$~\cite{Luo2017}, the ground-states should be two-fold degenerate, which are related with each other through shifting the vortex by half period along the leg direction. However, such degeneracy is lifted by the finite-size and finite-$n_\text{cutoff}$ effects.
In the strong interacting regime, the atom distribution in momentum space is broadened and the two wave packets at the two band minima would merge into one centered at $q=0$, therefore the system favors the M phase even when the lowest band has two minima. The current distributions of the interacting M and V phases are very similar as that for non-interacting case shown in Figs.~\ref{Noninteraction_phase}(a) and \ref{Noninteraction_phase}(b3).

Interestingly, a novel P phase emerges at intermediate interaction strength. The typical current and density distributions are shown in Fig.~\ref{Vortex II} for flux $\phi/\pi=1/4$. In this P phase, the mirror symmetry with respect to the middle leg is spontaneously broken and atoms start to occupy the dark \textcolor{red}{middle band with $n_-\neq0$}. Besides the mirror symmetry, the $Z_2$ exchange symmetry $|0  \rangle \!\! \leftrightarrow \!\! |+\rangle$ is also broken, leading to a four-fold degeneracy of the ground-states. \textcolor{red}{The symmetry breaking can be seen from the spin moments in Fig.~\ref{phase transition}(b) with $n_0\neq n_+$ and also the density distributions in Fig.~\ref{Vortex II} which indicate that $n_{j\uparrow}\neq n_{j\downarrow}$. Notice that the M and V phase preserve the mirror symmetry and $Z_2$ exchange symmetry with $n_0= n_+$ and $n_{\uparrow}= n_{\downarrow}$}. The P phase can be a combined population of the middle band with either left minimum [Figs.~\ref{Vortex II}(a) and \ref{Vortex II}(b)] or right minimum [Figs.~\ref{Vortex II}(c) and \ref{Vortex II}(d)] of the lowest band, and for either combination, the polarization $S_{j,z}=n_{j,\uparrow} - n_{j,\downarrow}$  is nonvanishing and can be either positive or negative (related by the mirror symmetry), which is determined by the relative phase between atoms on the two bands. For the two ground-states with dominate population on the left minimum of the lowest band, vortex-antivortex pairs are formed along the leg direction, leading to modulations of spin densities and currents with period $2\pi/\phi$. We want to mention that the order parmeters of the P phase in Figs.~\ref{phase transition}(a)-\ref{phase transition}(c) are obtained from the vortex ground-state. Moreover, due to the breaking mirror symmetry, the local spin-vector current $\sum_\sigma \sigma J_{ j,\sigma }^{\parallel }$ and spin-tensor current $\sum_\sigma \sigma^2 J_{j, \sigma }^{\parallel }$ are both finite in the P phase. This dark middle band has spin state orthogonal to the lowest band, so the superposition of the two bands will only induce spin orders without generating density modulation. Notice that both the P and V phases have a lower spin-spin interaction energy, the P phase has a higher single-particle energy but a lower density-density interaction energy compared to V phase, therefore a transition from V to P phase occurs as we increase the density-density interaction strength.

In Figs.~\ref{phase diagram}(a) and \ref{phase diagram}(b), we map out the phase diagrams in $\Omega \!\! - \!\! U_0$ plane with $\phi/\pi=1/4$  and in $\Omega \!\! - \!\! \phi$ plane with $U_0/t=2$, respectively. We see that the M phase area is enlarged by interactions.
Fig.~\ref{phase diagram}(a) clearly shows that the V and P phases are replaced eventually by the M phase as the interaction strength increases.  The system favors the P phase for weak inter-leg coupling $\Omega$ and small flux $\phi$, where dark middle band can be occupied more easily since the energy gap with the lowest band is small. For very small $\Omega$, the mixing between states $|0\rangle$ and $|+\rangle$ is also weak, so the density modulation and thereby the density-density interaction energy in the
V phase is weak, and the system favors the V phase instead of the P phase that has higher single-particle energy. Similarly, when the flux is large, density-density interaction energy for the V phase is small compared to the high single-particle energy in the P phase, and the system favors V over P phase, as shown in Fig.~\ref{phase diagram}(b). As $\Omega$ increases across some critical value, the density-density interaction energy for the V phase becomes stronger compared to the ferromagnetic spin-spin interaction energy, atoms will spontaneously populate around one of the two band minima with a single wave packet centered around $q\neq0$ and the system enters the M phase, further increasing $\Omega$ will drive the wave packet center to $q=0$.
For a small flux, the system may directly enter the M phase with one wave packet centered at $q=0$.
In this paper, we will not distinguish the two different types of M phase (centered around $q=0$ and $q\neq0$) since they have similar current distributions.

In obtaining the phase diagram, we have employed the finite-size scaling, leading to the critical points of phase transitions in the thermodynamic limit, which are almost the same as those of finite-size systems,
as shown in Fig.~\ref{phase diagram}(c). \textcolor{red}{The finite-size effects on the order parameters as well as the energy level spacing and gap are discussed in Appendix B.}
As we discussed previously, the V and P phases result from the ferromagnetic spin-spin interaction. For antiferromagnetic spin-spin interactions $U_2/U_0>0$, one only has the M phase. To see this, we plot the current order parameters as function of spin-dependent interaction strength $U_2$ in Fig.~\ref{phase diagram}(d), showing that the P phase only exists in the ferromagnetic region.
We want to emphasize that the Meissner currents induced by the gauge field persist even when the system enters the Mott insulator region, the red dashed line in Fig.~\ref{phase diagram}(a) shows the boundary between the superfluid and Mott insulator phases, which is obtained by calculating the chemical potential gap~\cite{gap1,gap2,gap3,gap4}.

\section{Discussion and Conclusion}
\label{Conclusions}
Experimentally, the spin moments can be measured
by population imaging after Stern-Gerlach separation~\cite{SDprl14,RamanRb}. The currents can be observed by site-resolved detection of the quench dynamics~\cite{Boseladder2}.
The current (motion of atoms) can also be proved by spin-selective  time-of-flight imaging of the lattice momentum distribution~\cite{Boseladder2, Mancini2015}
\begin{align}
n_{\sigma}(k) = \sum_{i,j}e^{ik(i-j)} \langle \hat{b}_{i \sigma}^{\dag}\hat{b}_{j \sigma} \rangle.
\end{align}
The current is related to the lattice momentum unbalance, and we can define the chirality of the atomic motion as
\begin{align}
\chi_{\sigma} = \frac{1}{L} \int_{0}^{\pi} h_{\sigma}(k) dk,
\end{align}
which shows similar behaviors as the current $J^{\parallel}_{\sigma}$~\cite{Mancini2015,Livi2016}, where $h_{\sigma}(k) = n_{\sigma}(k) - n_{\sigma}(-k)$ is the
asymmetry function.
Moreover, the entanglement entropy can be measured using
quantum interference of many-body twins of ultracold atoms in optical lattices~\cite%
{Islam2015}.

In summary, We studied the Meissner effects of interacting bosons {\color{red}in a one-dimensional} optical lattice with spin-tensor--momentum coupling. Using state-of-the-art density-matrix renormalization-group numerical method, we obtained the phase diagram with a rich variety of interesting phases, including the Meissner, vortex and polarized  phases.
The current distributions show spin-tensor properties and interesting vortex structures that are unique for the spin-tensor--momentum coupled system of high spin system, \textcolor{red}{which may find possible applications in atomic spintronic devices}. 
Our work reveals nontrivial phases and transport properties
resulting from the interplay between spin-tensors, lattice physics and interactions, and thus paves the
way for exploring novel many-body phenomena of interacting particles in nonuniform gauge fields.

\begin{figure}[b]
\centering
\includegraphics[width = 0.7\linewidth]{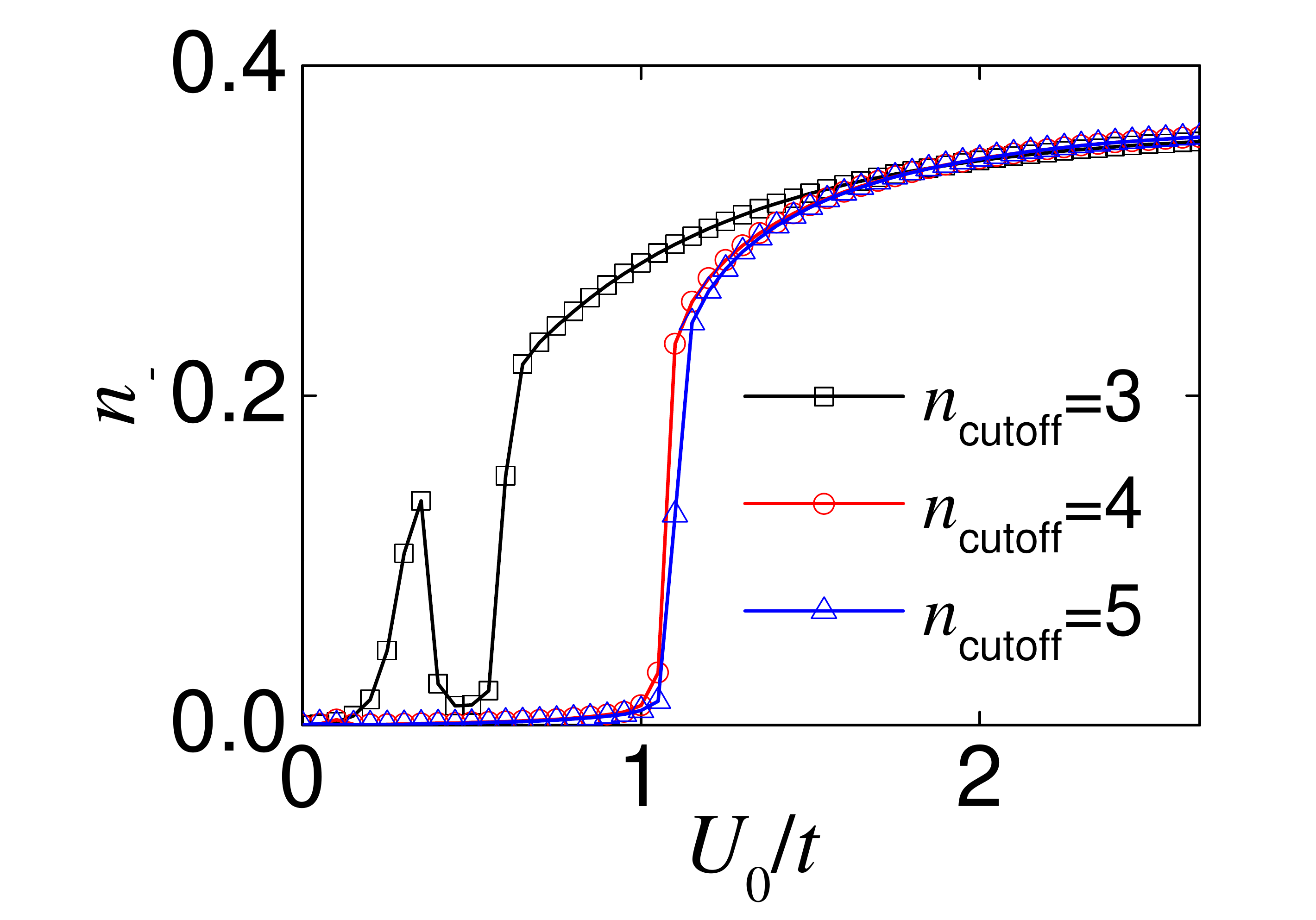} \hskip -0.0cm
\caption{Spin moment order $n_{-}$ as a function of interaction strength $U_0/t$, for several $n_\text{cutoff}$, with $\Omega/t=0.2$, $\phi/\pi=1/4$, $U_2=-0.2U_0$, $\rho = 1.0$ and $L=32$.}
\label{phase_transition_U_cutoff}
\end{figure}

\appendix
\textcolor{red}{\section{Effects of $n_\text{cutoff}$}}
As we discussed in the main text, the cutoff of single-site atom number $n_\text{cutoff}$ would modify the phase and phase boundary at the weak interacting region. 
A larger $n_\text{cutoff}$ leads to more accurate results, but the calculation becomes more computational expensive. 
In Fig.~\ref{phase_transition_U_cutoff}, we show the order parameter $n_-$ as a function of $U_0$ with different atom number cutoff. We see that, the results are nearly the same in the region $U_0/t>1.5$ for $n_\text{cutoff}=3,4,5$, while in the weak interacting region $U_0/t<1$, the results with $n_\text{cutoff}=3$
deviate significantly from that with $n_\text{cutoff}=4,5$.
This can be easily understood by noticing that,
for weaker interaction, the probability of high occupation is larger in the ground state.
On the other hand, we find that the results for $n_\text{cutoff}=4$ and $n_\text{cutoff}=5$ are nearly the same (both give the phase boundary around $U_0/t=1$), indicating that $n_\text{cutoff}=4$ is enough to determine the phase boundaries (notice that most of the phase boundaries are located in the region $U_0/t\geq1$ in the phase diagram). 

\begin{figure}[tb]
\centering
\includegraphics[width = 1.0\linewidth]{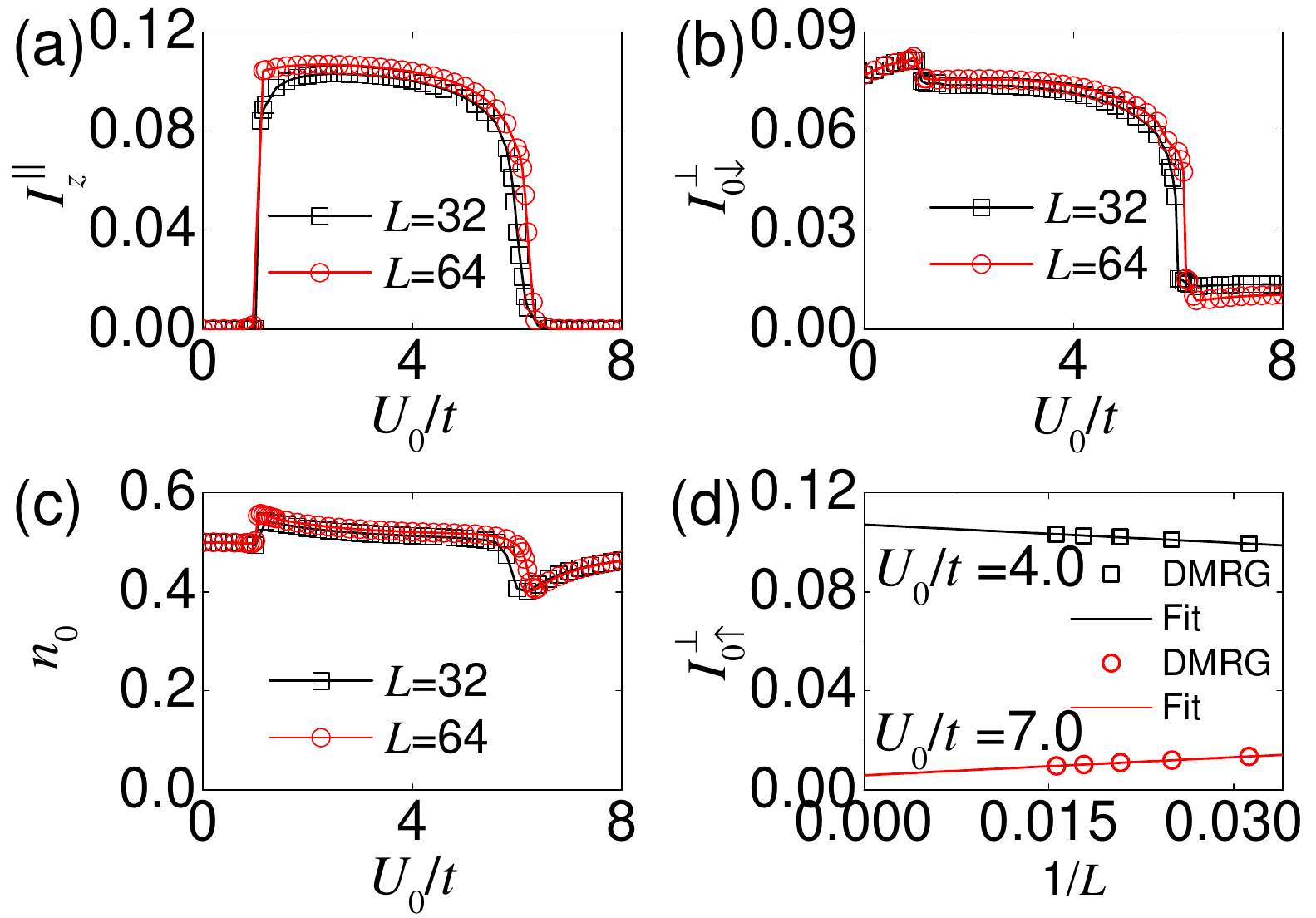} \hskip -0.0cm
\caption{(a) The current order parameter $I_{ z }^{\parallel }$, (b) the current order parameters $I_{0 \downarrow }^{\perp }$, and (c) the spin moment orders $n_{0}$ versus $U_0$ for $L=32$ and $64$.
(d) The finite-size scaling of the $I_{0 \uparrow }^{\perp}$ for $U_0/t=4.0$ and $7.0$.
In all subfigures, we have $\Omega/t=0.2$, $\phi/\pi=1/4$, $U_2=-0.2U_0$, and $\rho = 1.0$.}
\label{phase_transition_L32_L64}
\end{figure}

\begin{figure}[b]
\centering
\includegraphics[width = 1.0\linewidth]{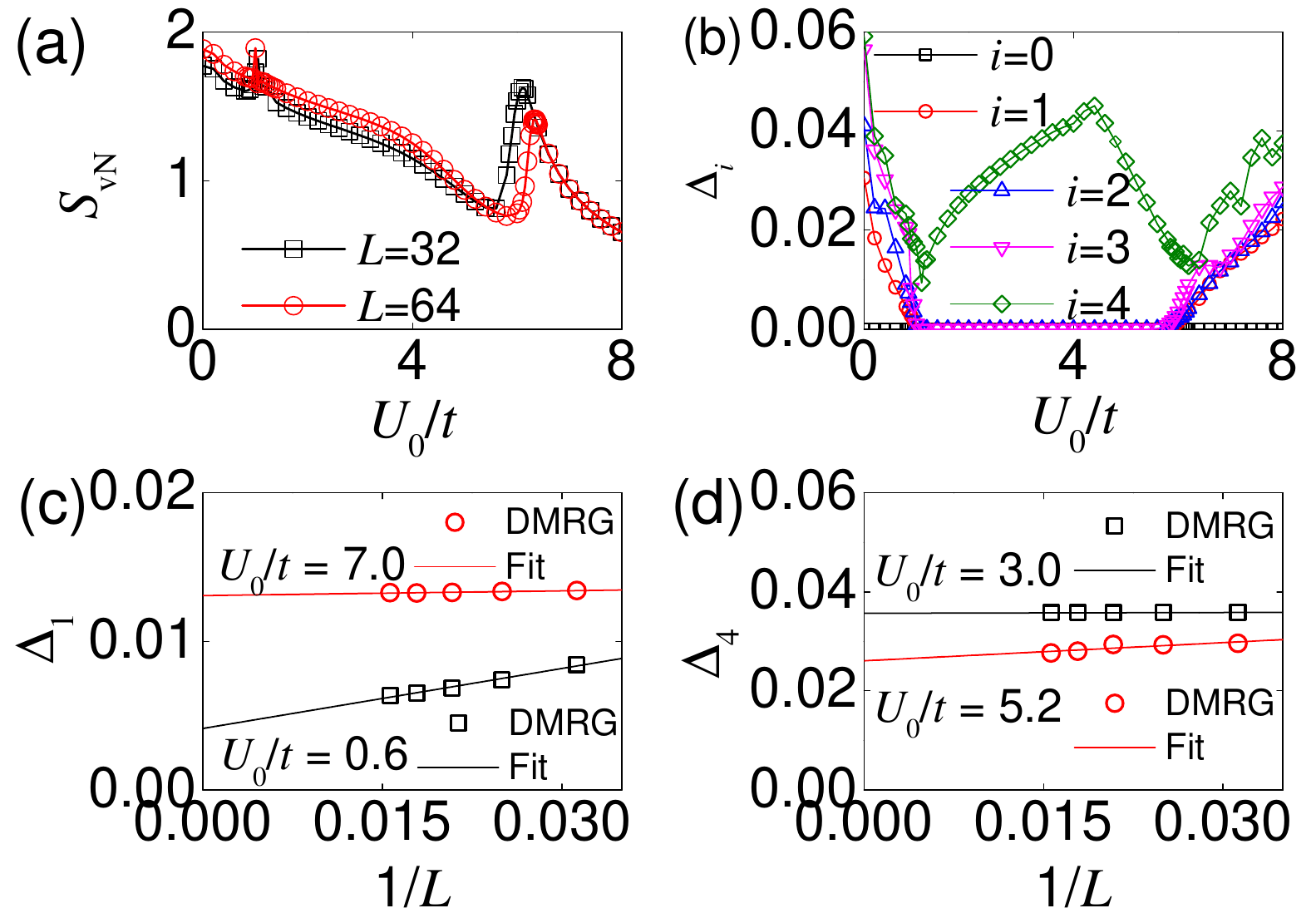} \hskip -0.0cm
\caption{(a) Entropy $S_{\text{vN}}$ versus $U_0/t$, for $L=32$ and $64$. (b) $\Delta_i$ versus $U_0/t$ for $L=32$. The finite-size scaling of $\Delta_i$ for (c) V phase with $U_0/t=0.6$ and M phase with $7.0$, and (d) P phase with $U_0/t=3.0$ and $5.2$. (d). In all subfigure, we have $\Omega/t=0.2$, $\phi/\pi=1/4$, $U_2=-0.2U_0$, $\rho = 1.0$. The unit of $\Delta_i$ is $t$.}
\label{phase_transition_S_Gap}
\end{figure}

\textcolor{red}{\section{Finite-size effects}}
Now we briefly discuss the finite-size effects on the order parameters.
As the system size increases, the order parameters may slightly modified (especially near the critical phase
boundaries), and thus the phase boundaries (determined by examining the spontaneous symmetry breaking) may shift slightly as we increase the size. The phase boundaries at the thermodynamic limit are obtained by finite-size scaling [see Fig.~\ref{phase diagram}(c)]. 
In Figs.~\ref{phase_transition_L32_L64}(a)-(c), we plot the order parameters as functions of $U_0$ for different system size $L=32$ and $L=64$, away from the phase boundaries, we see that the order parameters are almost the same for different sizes. In Fig.~\ref{phase_transition_L32_L64}(d), we plot the typical finite-size scaling of the order parameters and find that the thermodynamic value is very close to the finite-size value with $L=64$. We would like to mention that, the rung current in the M phase [see the red line in Fig.~\ref{phase_transition_L32_L64}(d)] decreases as the system size increase, this is because, the leg currents must form a closed loop at the boundary through rungs, such boundary-rung current lead to an averaged rung current scales as $1/L$. Numerically, we find that the averaged rung current $I_{0 s }^{\perp }$ does not vanish completely in the M phase (though it is extremely small). For such interaction driven M phase, though atoms occupy a single broadened wave packet in momentum space, the double-well structure of the lowest band may result in residual interference  that leads to the tiny rung current.

As we discussed in the main text, the von Neumann entropy $S_{\mathrm{vN}}$ as well as the low energy level spacing $\Delta_{i}$ can also signal the phase transitions. For different system sizes, the entropy $S_{\text{vN}}$ exhibits similar curves with peak positions always matching the phase boundaries given by the
order parameters (currents and spin moments), as shown in Fig.~\ref{phase_transition_S_Gap}(a) for $L=32$ and $64$. For the low energy level spacing, we have shown the results for $L=64$ in the main text, while the results for $L=32$ are similar [see Fig.~\ref{phase_transition_S_Gap}(b)], the four fold degeneracy is clearly seen in the P phase. To see how the energy gap scales with  system size, we plot the finite-size scaling results in Figs.~\ref{phase_transition_S_Gap}(c) and \ref{phase_transition_S_Gap}(d), notice that the energy gap corresponds to $\Delta_1$ ($\Delta_4$) for the V and M phase (P phase). 
The small gaps in the P and M phases change slightly with system size and remain finite (of the order of $\sim10^{-2}t$) in the thermodynamic limit, which vanish at their phase boundary. 
While for the V phase, the gap decreases significantly with increasing system size, reaching to a small value (of the order of $10^{-3}t$) in the thermodynamic limit. We also note that, in the V phase, higher excited-state energy gaps $\Delta_i$ ($i>1$) also decrease with increasing system size, the low energy levels are nearly equally spaced in the thermodynamic limit.
Since the V phase appears in the weak interaction region where the effects of cutoff $n_\text{cutoff}$ is more significant, we expect that the tiny gap in the V phase may be induced by the cutoff, 
and the V phase is probably a gapless phase. 

\textbf{\emph{Acknowledgments.---}}This work is supported by the National Key R\&D Program of China under Grant No. 2022YFA1404003, the National Natural Science Foundation of China (NSFC) under Grant No.~12004230, 12174233 and 12034012, the Research Project Supported by Shanxi Scholarship Council of China and Shanxi '1331KSC'. X.-W. Luo acknowledges support from the National Natural Science Foundation of China (Grant No. 12275203) and the USTC start-up funding.


\end{document}